# Electronic properties of Jahn-Teller and photoluminescence systems under pressure.

F. Rodríguez, University of Cantabria, DCITIMAC, Facultad de Ciencias, University of Cantabria, Santander, Spain, rodriguf@unican.es

**Summary**

Photoluminescence (PL) properties of materials containing transition-metal (TM) ions depend on a variety of structural factors such as electronic structure, site symmetry and neighbouring atoms. These factors play a crucial role for the occurrence PL *i.e.* the PL quantum yield. This work investigates different mechanisms leading to non-radiative de-excitation processes and whether they can be modified by applying high pressure. In particular, the interest is focussed on non-radiative-multiphonon relaxation in TM impurities. This non-radiative process is analysed in terms of the Dexter–Klick–Russell parameter, $\Lambda = 1/2\, \Delta E_S/E_{abs}$, which scales with the PL quantum efficiency. Depending on the material, the PL onset is favoured for $\Lambda < 0.1 - 0.3$. This work analyses the variation of $E_{abs}$ and $\Delta E_s$ for $Mn^{2+}$ and $Cr^{3+}$ as well as for the Jahn-Teller $Mn^{3+}$ in different coordination geometries as a function of pressure and the crystal host volume and the pressure.

**Introduction**

Photoluminescence (PL) properties of materials containing transition-metal (TM) ions depend on a variety of structural factors such as electronic structure, site symmetry and neighbouring atoms. These factors play a crucial role for the occurrence PL *i.e.* the PL quantum yield. This work investigates different mechanisms leading to non-radiative de-excitation processes and whether they can be modified by applying high pressure. In particular, the interest is focussed on non-radiative-multiphonon relaxation in TM impurities and PL quenching in concentrated materials, where exciton migration and subsequent energy transfer to non-PL centres take place. The former process is analysed in terms of the Dexter–Klick–Russell (DKR) criterion for the occurrence of PL through the parameter $\Lambda$, defined as $\Lambda = \frac{1}{2} \Delta E_S/E_{abs}$, where $\Delta E_s$ and $E_{abs}$ are the Stokes shift, and the absorption band energy, respectively (Dexter, 1955). The use of this parameter is advantageous since for a given system it has been probed that scales properly with the quantum efficiency, and its value can be predicted from Optical Absorption (OA) data alone (Marco, 1996). According to DKR criterion, the onset for PL is expressed as $\Lambda < 0.25$. In this model the PL is quenched by a competing non-radiative process whenever the intersection of the ground and excited state curves within a simple two-state configuration coordinate diagram lies below the energy reached in absorption in a vertical (Franck-Condon) transition (point A and C in Fig. 1). Empirically the PL quench is known to occur for $\Lambda > 0.25$ (Dexter, 1955; Bartram, 1975). In the case of TM systems, this criterion can change depending on the TM and its local structure but the PL onset for TM usually takes place for $\Lambda < 0.1 - 0.3$ (Marco, 1996). Following DKR criterion, we can be able to predict the evolution of PL efficiency of TM complexes with volume either in pressure experiments or in compound series through the electronic structure derived from the evolution of the OA spectra and the emission spectra in PL systems. In this paper we present and analyse the variation of $E_{abs}$ and $\Delta E_s$ with pressure and in compound series for $Mn^{2+}$ (Hernández, 2003; Marco, 1993, 1995, 1996; Rodriguez, 1986, 1991, 2003) and $Cr^{3+}$ (Ferguson, 1971; Güdel, 1978, 1986; Dolan 1986, 1992; Woods, 1993; Marco, 1995b; Wenger, 2003a,b) as well as the Jahn-Teller (JT) $Mn^{3+}$ (Aguado, 2003) in different coordination geometries.

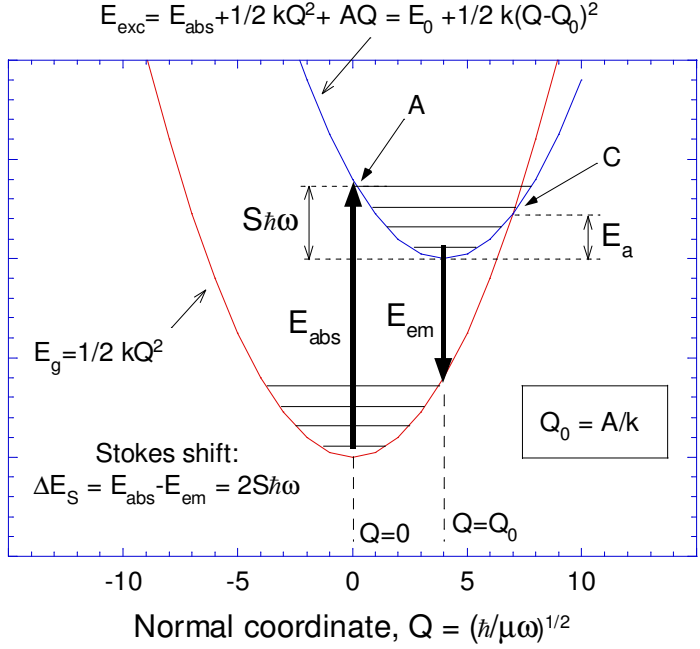

$E_{exc} = E_{abs} + 1/2\, kQ^2 + AQ = E_0 + 1/2\, k(Q-Q_0)^2$

A

C

$S\hbar\omega$

$E_a$

$E_g = 1/2\, kQ^2$

$E_{abs}$  $E_{em}$

$Q_0 = A/k$

Stokes shift:
$\Delta E_S = E_{abs} - E_{em} = 2S\hbar\omega$

$Q=0$  $Q=Q_0$

Normal coordinate, $Q = (\hbar/\mu\omega)^{1/2}$

**U**date configuration diagram for the excited and ground states. The DKR criterion for PL requires the intersection point C to be above the vertical excitation point B. The non-radiative deexcitation by multiphonon processes in a semiclassical description takes place by thermal activation through the activation energy $E_{act}$. Although harmonic models are unrealistic to predict actual activation energies, non-radiative process are blocked by increasing $E_{act}$. Parameters S, $\hbar\omega$, $E_{abs}$, $E_{am}$ and $E_0$ are the Huang-Rhys vibration energy, vertical absorption and emission energy, and zero-phonon line energy, respectively.

**Results and Discussion**

The temperature dependence of the PL lifetime (or the PL intensity) in competition with non-radiative deexcitation processes responsible for the PL quenching, can be in general described in terms of thermal activation terms as:

$$\tau^{-1} = \tau_0^{-1} + \tau_i^{-1}\mathrm{Coth}(\frac{\hbar\omega_i}{2kT}) + w_{nr}^i\, e^{-\frac{E_{act}}{kT}} \quad \text{and} \quad I_{PL}(T) = P\frac{\tau_0^{-1} + \tau_i^{-1}\mathrm{Coth}(\frac{\hbar\omega_i}{2kT})}{\tau_0^{-1} + \tau_i^{-1}\mathrm{Coth}(\frac{\hbar\omega_i}{2kT}) + w_{nr}^i\, e^{-\frac{E_{act}}{kT}}}$$

for the PL transition rate, $\tau^{-1}$, and intensity, $I_{PL}$, respectively. In these equations, $\tau_0^{-1}$ denotes the temperature-independent radiative probability, $\tau_i^{-1}\mathrm{Coth}(\frac{\hbar\omega_i}{2kT})$ is the vibronic assistance term associated with the i-th mode of energy $\hbar\omega_i$, and $w_{nr}^i\, e^{-\frac{E_{act}}{kT}}$ represents the transition rate of non-radiative process related to the i-th channel. These probabilities are additive for all channels involved in the non-radiative deexcitation decay like multiphonon relaxation, energy transfer to neighbouring ions, ionization through excitation to the conduction band, etc, and $E_{act}^i$ is the associated activation energy. The P parameter in $I_{PL}(T)$ accounts for the excitation power and the excited-to-ground state degeneracy ratio.

In the case of an isolated TM impurity whose PL quenching is mediated by non-radiative excited-state deexcitation by multiphonon processes, $E_{act}$ represents within a single coordinate configuration model of strongly coupled system ($S \gg 1$), the energy difference between the excited-ground state crossing point to the excited state minimum (Fig. 1). Within the harmonic approximation, the activation energy can be expressed as:

$$E_{act} = \frac{E_{em}^2}{4S\hbar\omega} \approx \frac{E_{abs}^2}{4S\hbar\omega} = \hbar\omega(\frac{p}{2S})^2$$

where $p = \frac{E_{abs}}{\hbar\omega}$ is the number of quanta required to bridge the excited state to the ground state energy difference for non-radiative relaxation. It is worth noting that the DKR parameter is proportional to the reciprocal of the activation energy thus justifying the adequacy of Λ as gauge for the onset of PL quenching (Λ > 0.25) and also for PL-efficiency through the activation energy. The smaller the Λ parameter, the more PL efficiency. In fact, Λ is defined as $\Lambda = \frac{1}{2}\frac{\Delta E_S}{E_{abs}} = \frac{1}{2}\frac{2S\hbar\omega}{p\hbar\omega} = \frac{S}{p}$, so that both parameters are related through the equation:

$$E_{act} = \hbar\omega(\frac{p}{2S})^2 = \frac{\hbar\omega}{(2\Lambda)^2}$$

In what follows our aim focuses in understanding how the volume influences the parameter Λ and whether a change of volume induced by hydrostatic pressure or in series of isomorphous compounds are comparable.

**Volume dependence of the Stokes shift and excitation energy.**
The electronic properties of TM placed either as impurity or as constituent of a material can be greatly explained on the basis of the molecular unit $MX_n$ formed by the metallic ion M and n ligands X, where n is 6 or 8 for octahedral and cubal coordinations, respectively (Sugano, 1970; Güdel, 1978, Lever, 1984). The role of the host crystal is mainly to exert a pressure on $MX_n$ yielding a given M-X bond distance, R, and coordination symmetry as a result. Although the $MX_n$ unit can be adequate to account for the electronic structure, it can fail however to describe the vibrational structure since force constants associated with the local vibrations can be influenced by the neighbouring atoms (Rodríguez, 1994; Barriuso, 2002). The Stokes shift associated with PL depends on the Huang-Rhys factor, $S_i$, of the coupled i-th vibration and its corresponding energy, $\hbar\omega_i$, through the equation $\Delta E_S = 2\sum_i S_i \hbar\omega_i$ where the summation extends to all coupled vibrations. In the case of $MX_n$ of cubic symmetry, the relevant coupled vibrations are the totally symmetric $a_{1g}$, and the JT $e_g$ modes in case of states of E or T symmetry were involved. Therefore R-dependence of $\Delta E_S$ requires knowledge on the relations between S and ω with R as well as between R and the crystal volume, V. The latter relation can be difficult to obtain since the local M-X distance around the impurity is in general different of the host distance (Rodriguez, 1986, 2003; Marco, 1994, 1995a, 1996). In addition, the variation of the M-X distance of the host crystal in elpasolites can scale differently than the cubic lattice parameter, $V^{1/3}$. These two facts make it tricky to predict volume dependences of the Stokes shift and electronic structure in impurity systems.

In terms of the electron-vibration coupling parameter, A, the Stokes shift S can be written as $\Delta E_S = \sum_i \frac{A_i^2}{\mu\omega_i^2}$ where μ is the reduced mass and $A_i = \left[\frac{\partial E_{exc}}{\partial Q_i}\right]_{Q_i = 0}$ for i = $a_{1g}$, $e_g$. In the case of PL-excited states whose energy is proportional to the crystal-field energy 10Dq (this is the usual situation for strongly coupled systems), then $A_{a_{1g}}$ depends on R as $A_{a_{1g}} \propto R^{-n-1}$, on the assumption that 10Dq varies with R as $10Dq \propto R^{-n}$. A similar dependence is derived for the JT coupling $A_{e_g} \propto R^{-n_e-1}$. Crystal-field theory, calculations and experiments provide n exponents between 4 and 6 for X = F and Cl (Zahner, 1961; Drickamer, 1973; Rodriguez, 1986; Woods, 1993; Barriuso, 2002). Therefore the $a_{1g}$ and $e_g$ terms of the Stokes shift depend on R as

$$\Delta E_S^{a_{1g}} = \frac{A_{a_{ia}}^2}{\mu\omega_{a_{1g}}^2} \propto R^{-2(n+1-3\gamma_a)} \quad \text{and}$$

$$\Delta E_S^{e_g} = \frac{A_{e_a}^2}{\mu\omega_{e_g}^2} \propto R^{-2(n_e+1-3\gamma_e)}. \quad (1)$$

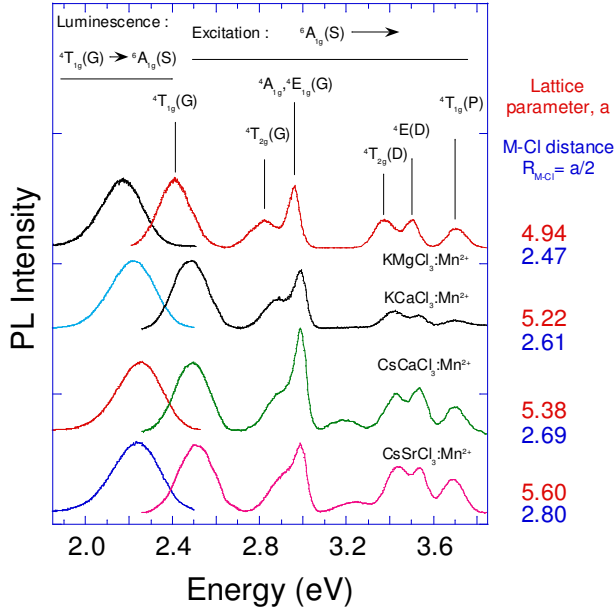
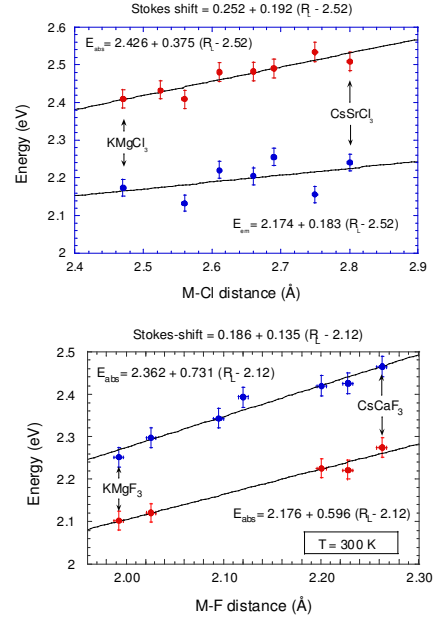

**Figure 2.** Left) Excitation and photoluminescence spectra of $Mn^{2+}$-doped $AMCl_3$ chloroperovskite series. The M-Cl distance for each crystal is given on the left side of the figure. Note the effect of decreasing the lattice parameter on the splitting and redshift exhibited by the three first bands. Right) Variation of the excitation and corresponding emission energies with the M-X distance for the two series $Mn^{2+}$-doped $AMX_3$ (X: Cl, F).

**TABLE I.** Structural and spectroscopic parameters of $Mn^{2+}$-doped $AMX_3$ perovskites and $Cr^{3+}$-doped $A_2BMX_6$ elpasolites (X: F, Cl). The Stokes shift and the BKR parameters are defined in the text. The Huang-Rhys factor, $S_e$, and the corresponding Jahn-Teller energy, $E_{JT} = S_e\hbar\omega_e$, for the $e_g$ mode are given for selected compounds. Units of the host M-X distance, R, and energies are given in Å and eV, respectively. References for perovskites: (Rodriguez, 1986, 1991; Marco, 1993, 1994, 1995; Hernández D., 1999; Hernández I., 2003a) and elpasolites: (Ferguson, 1971; Güdel, 1978; Dolan, 1986, 1992; Woods, 1993; Marco, 1995b; Wenger, 2001a,b).

| T = 300 K | Perovskites $AMX_3$: $Mn^{2+}$ | | | | | |
|---|---|---|---|---|---|---|
| System | R | 10Dq | $E_{ex}$ | $E_{em}$ | $\Delta E_S$ | $\Lambda$ | $S_e$ |
| $KMgF_3$: $Mn^{2+}$ | 1.993 | 1.045 | 2.252 | 2.102 | 0.150 | 0.033 | 1.35 |
| $KZnF_3$: $Mn^{2+}$ | 2.026 | 1.019 | 2.297 | 2.120 | 0.178 | 0.039 | 1.5 |
| $RbMnF_3$ | 2.120 | 0.929 | 2.393 | - | | | 1.6 |
| $CsCaF_3$: $Mn^{2+}$ | 2.262 | 0.854 | 2.465 | 2.275 | 0.190 | 0.038 | |
| $KMgCl_3$: $Mn^{2+}$ | 2.47 | 0.720 | 2.410 | 2.173 | 0.237 | 0.049 | |
| $NH_4MnCl_3$ | 2.525 | 0.690 | 2.430 | - | | | |
| $RbCaCl_3$: $Mn^{2+}$ | 2.66 | 0.675 | 2.482 | 2.205 | 0.277 | 0.056 | |
| $RbSrCl_3$: $Mn^{2+}$ | 2.75 | 0.627 | 2.534 | 2.156 | 0.379 | 0.075 | |

| T = 300 K | Elpasolites $A_2BMX_6$: $Cr^{3+}$ | | | | | |
|---|---|---|---|---|---|---|
| System | R | 10Dq=$E_{ex}$ | $E_{em}$ | $\Delta E_S$ | $\Lambda$ | $S_e$ | $E_{JT}$ |
| $K_2NaAlF_6$: $Cr^{3+}$ | 1.80 | 2.009 | 1.674 | 0.335 | 0.083 | | |
| $K_2NaGaF_6$: $Cr^{3+}$ | 1.87 | 1.984 | 1.676 | 0.308 | 0.078 | | |
| $Rb_2KGaF_6$: $Cr^{3+}$ | 1.89 | 1.978 | 1.686 | 0.292 | 0.074 | | |
| $K_2NaScF_6$: $Cr^{3+}$ | 1.98 | 1.965 | 1.637 | 0.328 | 0.083 | | |
| $Cs_2NaScCl_6$: $Cr^{3+}$ | 2.49 | 1.587 | 1.376 | 0.211 | 0.066 | 1.25 | 0.037 |
| $Cs_2NaInCl_6$: $Cr^{3+}$ | 2.51 | 1.612 | 1.363 | 0.249 | 0.077 | 1.3 | 0.038 |
| $Cs_2NaYCl_6$: $Cr^{3+}$ | 2.56 | 1.569 | 1.242 | 0.327 | 0.104 | 1.4 | 0.041 |

Note that within this model, the Stokes shift increases (or decreases) upon R reduction if $n+1 > 3\gamma$ (or $n+1 < 3\gamma$) where γ is the local Grüneisen parameter of the coupled vibration, i.e. γ is defined with respect to the actual M-X impurity distance. In this model γ does not scale with the crystal volume but it does with the local R distance, and consequently may differ from each other. Taking common values of n = 5 and γ ~ 1 – 2, the model predicts an increase of the Stokes shift with decreasing R thus with increasing pressure. However, it must be noted that values γ > 2 has been calculated for $CrF_6^{3-}$ and $MnF_6^{4-}$ (Woods, 1993; Barriuso, 2002) thus suggesting that Grüneisen parameters of local vibrations may be significantly different depending on whether frequency variations are referred to the volume of the $MX_n$ complex (γ > 2) or the crystal volume (γ ~ 1) (Barriuso, 2002). Indeed both descriptions can be concealed if we consider local relaxation effects around the impurity (Rodriguez, 1986). This fact shows that estimating pressure behaviours of the Stokes shift, hence the PL efficiency, is a difficult task from the present model.

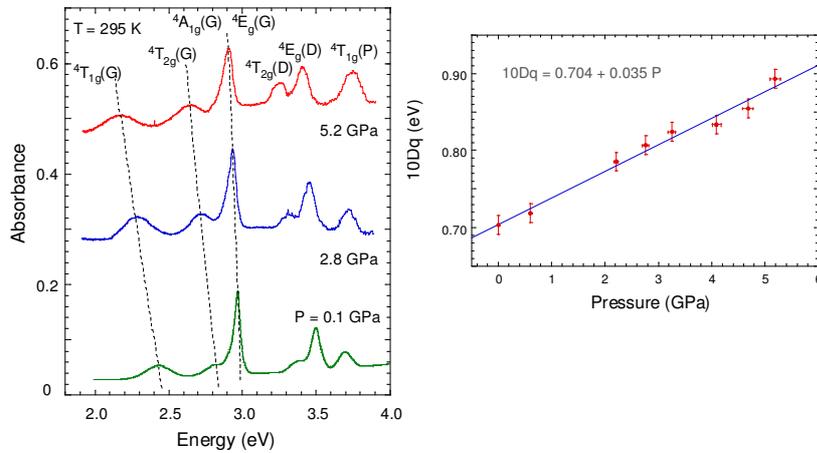

**Figure 3.** Left) Variation of the optical absorption spectrum of $NH_4MnCl_3$ with pressure. The band shifts are similar to those of Fig. 2 upon reduction R. Right) Variation of the crystal-field energy, 10Dq, with pressure. The crystal field increases 35 meV $GPa^{-1}$.

For this purpose we firstly study the variation of the electronic structure, PL and Stokes shift by optical spectroscopy in two compound series: $AMX_3$ (X: Cl, F) perovskites doped with $Mn^{2+}$ and $A_2BMX_6$ elpasolites doped with $Cr^{3+}$, and as a function of pressure. The selected $Mn^{2+}$ and $Cr^{3+}$ ions are reliable for PL studies since they exhibit an unique PL emission associated with the first excited state: $^4T_{1g}$ for high-spin $Mn^{2+}$, and $^4T_2$ or $^2E$ for $Cr^{3+}$, depending on the crystal-field strength to be below or above the $^4T_2$-$^2E$ excited-state crossover point, respectively (Sugano, 1970). Secondly, the JT energy of $Mn^{3+}$ in the fluoride series $AMnF_4$ (A: Na, Tl, Cs) is analysed as an illustrative example to determine the variation of the JT energy as a function of pressure (Aguado, 2003). Finally, the pressure-induced transformation from non-PL to PL in the fluorites $MF_2$: $Mn^{2+}$ (M. Ca, Ba) and in $MnF_2$ (rutile structure) are analysed on the basis of present models.

**Stokes shift and excitation energy in $Mn^{2+}$-doped $AMX_3$ and $Cr^{3+}$-doped $A_2BMX_6$ (X: Cl, F).**
Figure 2 shows the variation of the excitation and emission spectra of $Mn^{2+}$-doped $AMCl_3$ chloroperovskites. These crystals are attractive for accommodating $Mn^{2+}$ in octahedral sites. The spectra are characteristic of $MnCl_6^{4-}$ whose crystal-field 10Dq and Racah B and C parameters are given elsewhere (Marco, 1995a). The spectra are shown from $KMgCl_3$ to $CsSrCl_3$ in order of increasing lattice parameter. Note that the Stokes shift increases with the crystal volume, i.e. decrease with pressure. A similar behaviour is also observed in fluoroperovskites (Rodriguez, 1986; Marco, 1994) as Fig. 2 shows. The excitation energies shift upon crystal volume reduction according to expectations on the basis of the Tanabe-Sugano diagram for an increase of 10Dq. Table I collects the structural and spectroscopic parameters of the two series. Note that the measured 10Dq values scale with R as $R^{-n}$ but n ≈ 3 (n ≠ 5). In fact, the value (10Dq)/ ΔR = -0.24 eVÅ$^{-1}$ yields an exponent $n = -\frac{\Delta(10Dq)}{\Delta R}\frac{R}{10Dq} = 0.24\frac{2.52}{0.7} \approx 1$. The deviation of the exponent n with respect the expected value n =5 is due to a strong local relaxation effects around the impurity. The actual Mn-Cl (or Mn-F in $AMF_3$) distance is longer than the host M-X distance depending on whether $R_{M-Cl}$ < 2.52 Å (or $R_{M-F}$ < 2.12 Å) or viceversa (Marco, 1993, 1994, 1995a). Similar band shift behaviour is observed under hydrostatic pressure in $NH_4MnCl_3$ (Fig. 3)(Hernandez, 1999). The corresponding variation of 10Dq with pressure is also shown Fig. 3. It is worthwhile noting that the Stokes shift decreases with pressure in the two series of fluoro- and chloroperovskites, hence the corresponding Huang-Rhys parameter S must also decrease with pressure since the vibrational frequency for stretching modes increases with pressure. The present finding has been clearly shown for the JT $e_g$ vibration whose contribution to the Stokes shift, also called JT energy, $E_{JT} = S_e \hbar \omega_e$ is known for several $Mn^{2+}$ and $Cr^{3+}$ systems through the reduction of the spin-orbit splitting of the respective $^4T_{1g}$ and $^4T_{2g}$ states by the Ham effect (Rodriguez, 1991; Güdel, 1978; G Wenger, 2001a,b). Table I collects the Huang-Rhys parameters as a function of the M-X distance of the host crystal. Note that $\Delta E_S$ in $Cr^{3+}$-doped fluoroelpasolites does not follow the same trend observed in chloroelpasolites and the two series of $Mn^{2+}$. Recent pressure experiments carried out in pure fluoroelpasolites $A_2BCrF_6$ (Hernández, 2005) suggest that the increase of $\Delta E_S$ with R is due to the weaker R-dependence of the first excitation peak $^4T_{2g}$ (10Dq) likely due to the Fanno resonance present in $Cr^{3+}$ in fluoroperovskite. At variance with chloroelpasolites (10Dq=1.6 eV), the crystal field energy for $CrF_6$ (10Dq = 2.0 eV) lies near the excited-state $^4T_2$-$^2E$ crossover point thus leading to different variations of the first excitation band with pressure. The bandshift variation is intermediate between the 10Dq-dependent $^4T_2$ and the 10Dq-independent $^2E$ thus providing values n < 3 (Hernández, 2005). Taking values of γ = 1.5 for the $a_{1g}$ and $e_g$, respectively, (Dolan, 1986; Wenger, 2001a) it turns out from Eq. (1) that $\Delta E_S$ should decrease with R if n ~ 3. Although Stokes shift calculations indicate that $\Delta E_S$ increases with R in $CrF_6$ (Dolan, 1986), the experimental values of Table I suggest an opposite behaviour. The reason of such a discrepancy lies on the fact that calculations of the $^4T_2$ energy provide values of n near 5 and according to Eq. (1) to positive variations of $\Delta E_S$ with R. On This situation is not observed in $CrCl_6$ since the $^2E$ and $^2T_{1g}$ excited states lie far above the $^4T_{2g}$ PL state yielding bandshift dependences as $R^{-6}$ (Dolan, 1986) thus leading to negative $\delta\Delta E_S/\delta R$ ratios, i.e. negative $\delta\Delta E_S/\delta P$ ratio. The data of spectroscopic and structural data given in Table I illustrates that behaviour. Interestingly, the Stokes shift in $CrCl_6$ reduces with hydrostatic pressure such as it was shown in $Cr^{3+}$-doped $Cs_2NaScCl_6$ by high pressure and low temperature spectroscopy (Wenger, 2001b). Note also that the Huang-Rhys parameter increases with increasing the M-X distance for all investigated systems. This result is noteworthy since aside the temperature dependence of the non-radiative deexcitation rate, the Huang-Rhys factor S plays a crucial role in the multiphonon relaxation probability at 0 K (tunnelling process) which is proportional to $\tau_{nr}^{-1} = P_{nr} e^{-S} \frac{S^p}{p!}$ (Bartram, 1986). It must be emphasized that the variation of $E_{JT}$ and S with R has been obtained indirectly through the reduction of the spin-orbit splitting of the $^4T_{2g}$ state in $Cr^{3+}$ (or $^4T_{1g}$ in $Mn^{2+}$) by the Ham effect (Rodriguez, 1991; Wenger, 2001a). The knowledge on how the JT energy depends with pressure on a given system is not easy to achieve directly through experiments. For this reason the study of JT systems is advantageous for such purpose (Rodriguez, 1994).

**Variation of the Jahn-Teller energy with pressure in $NaMnF_4$ and $CsMnF_4$.**

$Mn^{3+}$ is a JT ion whose electronic structure provides direct information on the JT energy associated with the $E_g$ and $T_{2g}$ octahedral states. In fluorides, the OA bands appear well resolved allowing a pressure dependence study (Rodriguez, 1994; Aguado, 2003). Figure 4 shows the variation of the OA spectrum of $NaMnF_4$ with pressure. A comprehensive band assignment is given elsewhere (Aguado, 2003). The three broadband energies provide the splitting pattern associated with the JT distortion such as is shown in the energy level diagram of Fig. 4. Note that $E_1$ and $E_3$-$E_2$ are the splitting of the parent octahedral $E_g$ and $T_{2g}$ states, $\Delta_e$ and $\Delta_t$, respectively. The former splitting is related to the JT energy as $E_1 = \Delta_e = 4E_{JT}$ so that its variation with pressure can be obtained from the pressure-induced shift of $E_1$. The results of Fig. 4 indicate that the JT energy, $E_{JT} = E_1/4 = 0.4$ eV at ambient pressure experiences a small blueshift with pressure: $\frac{\partial E_1}{\partial P} = 3$ meVGPa$^{-1}$ in $NaMnF_4$. Nevertheless, the JT shift strongly depends on the structural change undergone by the MnF$_6$ unit under pressure in the lattice. In fact, the pressure shift in CsMnF$_4$ in the tetragonal phase is $\frac{\partial E_1}{\partial P} = -4$ meV·GPa$^{-1}$ which is opposite to the pressure shift of NaMnF$_4$ (Aguado, 2005). The different shift is associated with the higher reduction of the JT distortion attained in pressure experiments. The bigger size of Cs$^+$ in comparison to Na$^+$ impedes tilting phenomena at expenses of a higher reduction of the JT distortion thus yielding reduction of the JT energy (Aguado, 2005). Pressure induces a reduction of the JT distortion in both systems and minor variations of the JT energy. The present results agree with previous data on Cr$^{3+}$ and Mn$^{2+}$ on the basis of volume variations through compound series, but constitutes a first prove of pressure-induce JT release in MnF$_6$.

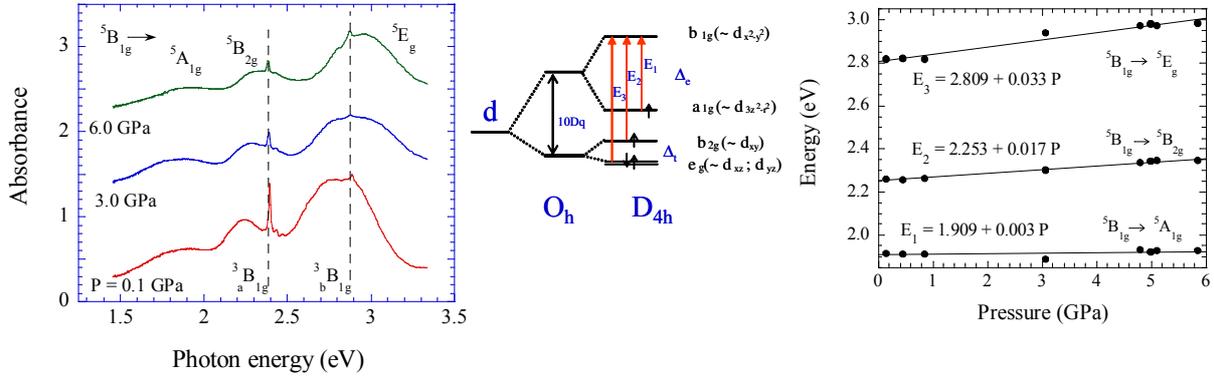

**FIGURE 4**. Variation of the optical absorption spectrum of NaMnF$_4$ with pressure at room temperature. Vertical lines indicate the position of the $^5B_{1g} \rightarrow ^3B_{1g}$ spin-flip transitions. The level diagram within a D$_{4h}$ symmetry, the electronic transitions and the pressure shifts are also included. The band assignment corresponds to $^5B_{1g} \rightarrow ^5A_{1g}$ (E$_1$), $^5B_{2g}$ (E$_2$), and $^5E_g$ (E$_3$),

**Pressure-induced PL in MF$_2$ (M: Ca, Ba, Mn)**

Mn$^{2+}$-doped MF$_2$ (M: Ca, Sr, Ba) fluorites are an illustrative example of interplay between PL and non-radiative processes leading to PL quenching (Diaz, 1999; Rodriguez, 2003). Figure 5 shows the variation of the Mn$^{2+}$ PL lifetime as a function of temperature along the Mn$^{2+}$-doped Ca$_{1-x}$Sr$_x$F$_2$ series. The results point out that the activation energy associated with non-radiative processes progressively decreases with the lattice parameter, or the M-F distance, R, as indicated in Table II. The PL trend on passing from CaF$_2$ to SrF$_2$ indicates that the BKR parameter and the activation energy strongly vary with R: E$_{act}$ changes from 1.02 to 0.10 eV whereas Λ increases from 0.048 to 0.068 according to the increase of R from 2.37 to 2.51 Å. From the variation of E$_{act}$ it turns out that Mn$^{2+}$-doped BaF$_2$ (R = 2.69 Å) exhibit no PL associated with MnF$_8$ impurities (Diaz, 1999; Hernandez, 2003b). The absence of PL in Mn$^{2+}$-doped SrF$_2$ and BaF$_2$ at room temperature relates to the small activation energies in comparison to CaF$_2$ (PL efficiency ~ 1). The loss of PL efficiency along the series correlates with an increase of Λ from 0.048 in CaF$_2$ to 0.065 in SrF$_2$, reflecting the increase of Stokes shift from 0.27 to 0.39 eV. It is worth noting that CrCl$_6$, CrF$_6$ and MnCl$_6$ systems exhibit similar or even bigger Λ values but are Pl with quantum efficiencies near to 1. This specific behaviour of MnF$_8$ is likely related to the fluorite structure which provides cubal coordination for Mn$^{2+}$. Enhancement of second order and anharmonic effects in this coordination can probably account for the distinct behaviour.

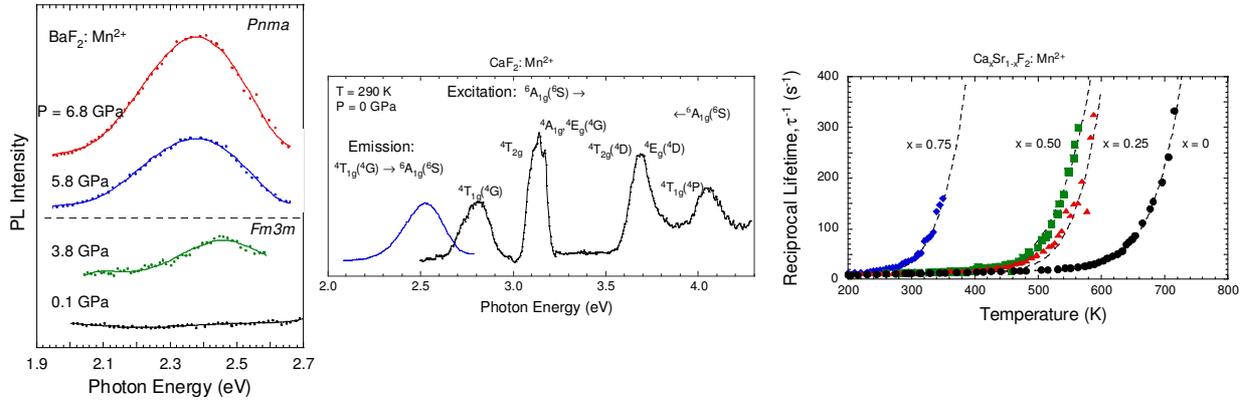

**FIGURE 5**. Variation of the PL spectrum of BaF$_2$: Mn$^{2+}$ with pressure at room temperature. Note the occurrence of PL above 3.8 GPa. The excitation and PL spectra of the PL CaF$_2$: Mn$^{2+}$ with the peak assignment is shown in the centre. The thermal dependence of the reciprocal lifetime for Ca$_x$Sr$_{1-x}$F$_2$: Mn$^{2+}$ illustrates the variation of the PL-quenching temperature with the content of strontium. The corresponding activation energies are given in Table II.

**TABLE II**. Structural and spectroscopic parameters of Mn$^{2+}$-doped AMX$_3$ perovskites and MF$_2$: Mn$^{2+}$ fluorites. The errors are 0.012 and 0.02 eV for energy in perovskites and fluorites, respectively. The activation energy for non-radiative deexcitation, E$_{act}$, has an error of 0.10 eV. The host M-X distance, R, is given in Å. References for perovskites: (Rodriguez, 1986; Marco, 1993, 1994, 1996) and fluorites: (Marco, 1996; Diaz, 1999; Hernandez, 2003; Rodriguez, 2003).

| T = 300 K | Perovskites AMX$_3$: Mn$^{2+}$ and Fluorites MF$_2$: Mn$^{2+}$ | | | | | | |
|---|---|---|---|---|---|---|---|
| System | R | 10Dq | E$_{ex}$ | E$_{em}$ | ΔE$_S$ | Λ | E$_{act}$ |
| KMgF$_3$: Mn$^{2+}$ | 1.99 | 1.045 | 2.252 | 2.102 | 0.15 | 0.033 | - |
| CsCaF$_3$: Mn$^{2+}$ | 2.26 | 0.854 | 2.465 | 2.275 | 0.19 | 0.038 | - |
| CaF$_2$: Mn$^{2+}$ | 2.37 | 0.51 | 2.80 | 2.53 | 0.27 | 0.048 | 1.02 |
| SrF$_2$: Mn$^{2+}$ | 2.51 | 0.47 | 2.87 | 2.48 | 0.39 | 0.068 | 0.10 |
| BaF$_2$: Mn$^{2+}$ | 2.69 | - | - | non-PL | - | - | 0 |

According to previous results, the application of pressure transforms the non-PL Mn$^{2+}$-doped BaF$_2$ into PL material (Fig. 5). Pressure-induced PL at room temperature was also obtained in the non-PL Mn$^{2+}$-doped SrF$_2$ (Hernández, 2003). These noteworthy results clearly demonstrate the generally accepted empirically-based rule for impurity systems by which PL is more favoured in smaller host sites.

Following this idea we intend to induce PL at room temperature in MnF$_2$ (rutile-type structure) by applying pressure. Preliminary results indicate that the structural changes induced by pressure in MnF$_2$ can efficiently enhance PL processes (Hernandez, 2005). Further work is currently in progress.

**Conclusions**

The present results reveal that pressure is an efficient tool to either enhance or induce PL in TM-doped inorganic materials. The variation of the Stokes shift with the crystal volume and pressure seems to be an adequate target for the search of efficient PL materials. In conclusion, the Huang-Rhys factor associated with the a$_{1g}$ and e$_g$ modes and corresponding energy, Sℏω, increase with the crystal volume or decrease with the pressure in MnCl$_6$, MnF$_6$, and CrCl$_6$. The only exception to this behaviour is the CrF$_6$ system as a result of the Fanno resonance in the PL state. Therefore TM PL in inorganic materials can be enhanced either by reduction of the host site volume or by applying pressure. Based on this idea it was possible to transform the non-PL Mn$^{2+}$-doped BaF$_2$ into PL by applying pressure. Structural correlations suggest that the cubal coordination provides smaller activation energies for non-radiative deexcitation than octahedral coordination, thus justifying the PL quench exhibited by $Mn^{2+}$-doped $BaF_2$ and $Mn^{2+}$-doped $SrF_2$ at room temperature even if the BKR parameter was smaller than in other PL systems of $Cr^{3+}$ and $Mn^{2+}$ with octahedral symmetry.

**Acknowledgments**
The author wishes to thank Drs. C. Marco de Lucas, R.E. Gutiérrez, F. Aguado and I. Hernandez for collaboration and fruitful discussions. Financial support from the Spanish MEyC (Project No MAT2005-00099) is acknowledged.